\documentclass[pra,twocolumn,showpacs,preprintnumbers]{revtex4}
\usepackage{graphicx}
\usepackage{amsfonts}
\usepackage{amssymb}
\usepackage{amsmath}
\makeindex

\newcommand{\be}{\begin{equation}}
\newcommand{\ee}{\end{equation}}
\newcommand{\bea}{\begin{eqnarray}}
\newcommand{\eea}{\end{eqnarray}}
\newcommand{\Eq}[1]{Eq.\,(\ref{#1})}% \Eq{abc}
\newcommand{\Fig}[1]{Fig.\,\ref{#1}}% \Fig{fig:abc}
\newcommand{\Sec}[1]{Sec.\,\ref{#1}}% \Sec{sec:abc} sic!byc konsekewntnym \label{sec:xx} \Sec{sec:xx}
\newcommand{\GFHk}{\hat{\mathbf{G}}^H_k}
\newcommand{\GFEk}{\hat{\mathbf{G}}^E_k}
\newcommand{\GFHK}{\hat{\mathbf{G}}^H_k}
\newcommand{\GFEK}{\hat{\mathbf{G}}^E_k}
\newcommand{\GFk}{\hat{\mathbf{G}}_k}

\newcommand{\br}{\mathbf{r}}

\newcommand{\En}{\mathbf{E}_n}
\newcommand{\Enb}{\mathbf{E}_{\bar{n}}}
\newcommand{\Hnb}{\mathbf{H}_{\bar{n}}}
\newcommand{\Hmb}{\mathbf{H}_{\bar{m}}}
\newcommand{\Hn}{\mathbf{H}_n}

\newcommand{\Emb}{\mathbf{E}_{\bar{m}}}
\newcommand{\Enpb}{\mathbf{E}_{\bar{n}'}}

\newcommand{\bE}{\bf\mathbf{E}}
\newcommand{\bH}{\bf\mathbf{H}}
\newcommand{\bJ}{\bf\mathbf{J}}
\newcommand{\Epn}{\mbox{\boldmath ${\cal E}$}\hspace*{-1.5pt}_\nu}
\newcommand{\heps}{\hat{\boldsymbol{\varepsilon}}}

\newcommand{\brd}{\mathbf{\dot{r}}}
\newcommand{\brho}{\mbox{\boldmath ${\rho}$}}

\newcommand{\epshat}{\hat{\boldsymbol{\varepsilon}}}

\begin{document}
%\pagewiselinenumbers
\title{Resonant state expansion for Transverse Electric modes of two-dimensional open optical systems}
\author{ M.\,B. Doost}
\affiliation{Independent Researcher,
United Kingdom}
\begin{abstract}
The resonant state expansion (RSE), a rigorous perturbative method in electrodynamics, 
is formulated for Transverse Electrodynamic modes of an effectively $2$-dimensional system. 
The RSE is a perturbation theory based on the Lippmann Schwinger Green's 
function equation and requires knowledge of the Green's function of 
the unperturbed system constructed from Resonant-states.
I use the analytic Green's 
function for the magnetic field to  normalize the modes appearing in the corresponding spectral 
Green's function. This use of the residue and cut of the  analytic Green's function is a solution 
to the failure of the flux volume integral method of normalization for continuum  states, a problem which is discussed
in detail in this manuscript. In brief, the flux 
volume integral fails for continuum states because they are not true resonance. An analytic  
relation between normalized magnetic and electric modes is developed as part of the solution to these difficulties.
The complex eigenfrequencies of modes 
are calculated using the RSE for the case of a homogeneous perturbation.
 
\end{abstract}
\pacs{03.50.De, 42.25.-p, 03.65.Nk}
\date{\today}
\maketitle
\section{Introduction}

Dielectric microcavities have attracted significant interest since 1989 when they were found to support whispering gallery modes (WGMs) \cite{Braginsky89}, 
particularly with regards to the 
development of hybrid optoelectronic circuits \cite{Vahala03}. Perturbed microcavities have proved to be the most promising direction of research
since their emissions are not isotropic \cite{Nockel97,Gmachl98,Lee02,Harayama03,Chern03,Kurdoglyan04,Wiersig08}. Other uses of $2$-dimensional
optical microcavities include miniature lasing devices \cite{Vahala44}.

Recently there has been developed the Resonant-state-expansion (RSE) a theory for calculating the resonances of open optical systems 
\cite{ArmitagePRA14,DoostPRA13,DoostPRA14,DoostPRA12,MuljarovEPL10,Doost_Muljarov}. That the RSE is suitable 
for use with the RSE Born approximation, to calculate the far field scattering, has been demonstrated in Ref.\cite{Doost15,Doost16,Ge14}.
The RSE Born approximation is a rigorous perturbation method for electrodynamic scattering, which becomes the exact solution of Maxwell's
wave equations in the far field, in the limit of an infinite number of resonance being taken into account in the Born approximation expansion.

In this paper I report extending the RSE to the Transverse Electric (TE) modes of 
effectively two-dimensional (2D) systems (i.e. 3D systems translational invariant in one
dimension) which are {\it not reducible} to effective 1D systems. I only treat systems with zero wavevector along the translational 
invariant direction at present. I use
a dielectric cylinder with uniform dielectric constant in vacuum as unperturbed system and
calculate perturbed Resonant-states (RSs) for homogeneous perturbations in order to demonstrate the method.

The treatment in this manuscript is restricted to $2$-dimensional systems, however for the modes of a cavity with mode wavelength only
a small fraction of the thickness of the cavity it is possible to make the mathematical approximation to an effectively $2$-dimensional system
\cite{Smotrova05, Lebental07}. A consequence of this approximation for the mathematical treatment is that the refractive index $n_r$ has to be replace with an 
appropriately calculated $n_{eff}$ which takes into account the thickness of the resonator, for the detailed mathematics, see Ref.\cite{Smotrova05, Lebental07}.

In this manuscript dispersion in the refractive index is not treated directly, however it is straight-forward to add dispersion using the method from Ref.\cite{Doost_Muljarov},
which I developed in collaboration with E. A. Muljarov, based on my successful initial independent work incorporating dispersion into the RSE.

Interestingly the RSE is a near identical translation to electrodynamics 
of a much earlier theory from Quantum Mechanics by More, Gerjuoy, Bang, Gareev, Gizzatkulov 
and Goncharov Ref.\cite{More73,Bang78}. The only difference between the two approaches is the choice of RS normalisation method.
The RSE \cite{More73,Bang78} is a perturbation theory based on the Lippmann Schwinger Green's function equation and requires knowledge of the Green's 
function (GF) of 
the unperturbed system constructed from RSs.
I was able
to show in Ref.\cite{Doost15} that the general normalisation of RSs which I derived in Ref.\cite{DoostPRA14} is the most numerically 
stable available normalisation method. The general normalisation which I derived in Ref.\cite{DoostPRA14} is based on a prototype normalisation which appeared in 
Ref.\cite{MuljarovEPL10}.

One difficulty in formulating an RSE for 2D systems,
is the presence of a one-dimensional continuum in the manifold of RSs. This continuum is specific to 2D systems and is required for the completeness of the basis 
and thus for the accuracy of the RSE applied in 2D, as discussed when
similar observations were previously made for Transverse Magnetic (TM) modes of a microcylinder in my Ref.\cite{DoostPRA13}.

In Ref.\cite{DoostPRA13,DoostPRA14} it was shown that the continuum states in the RSE cannot be normalised by the flux volume normalisation,
\bea\label{HELLOGOODBYE}
1+\delta_{k_n,0}&=&\int_A\En(\brd)\cdot\dfrac{\partial \omega^2\heps(\brd,\omega)}{\partial \omega^2}\bigg|_{\omega=\omega_n}\!\!\En(\brd)d{\brd}\\
&&+\lim_{k\rightarrow k_n}\oint_{L_A} \dfrac{\En\cdot\nabla\bE-\bE\cdot\nabla\En}{k^2-k^2_n} d{\bf L}\nonumber
\eea
I will discuss the reason for this in more detail in \Sec{sec:Maxwell2D}. Instead of using \Eq{HELLOGOODBYE}, it is necessary to make a comparison
of the residue and contour integral between the spectral GF
\be \GFk(\br,\br')=\lefteqn{\sum_n}\int \,\frac{\En(\br)\otimes\En(\br')}{2 k(k-k_n)}\approx\sum_{\bar{n}}\frac{\phi_{\bar{n}}\Enb(\br)\otimes\Enb(\br')}{2 k(k-k_{\bar{n}})}
\ee
and the analytically derived GF. The electrodynamic wave function $\Enb$ is the solution to 
\be\label{RS22}
\nabla\times\nabla\times\Enb(\br)=k^2_{\bar{n}}\heps(\br)\Enb(\br)
\ee
with outgoing boundary conditions (BCs) and $\GFk(\br,\br')$ is the corresponding GF for \Eq{RS22}.

Unfortunately it is not possible to derive the analytic GF for the electric $\bE$-field since for $2$-dimensional TE modes 
of an open system Maxwell's wave equation is not reducible to one dimension whilst in the electric field description. However the magnetic $\bH$-field GF is 
available and in Appendix A I use physical arguments to construct an analytic formula linking normalised magnetic modes $\Hnb$ to normalised electric modes $\Enb$.
The analytic Green's function for the homogeneous $2$-dimensional microcylinder is given in Appendix B.

This manuscript is organised as follows
\Sec{sec:AAAA1} outlines the derivation of the RSE perturbation theory,
\Sec{sec:Maxwell2D} explains the special provisions made for the normalisation of the $\Enb$ modes,
\Sec{sec:UPBF2D} gives the basis modes of the $2$-dimensional RSE, 
\Sec{sec:Hom} contains the numerical validation of the rigorous analytics provided in this manuscript, 
Appendix A and B derive the normalisation of the TE $\Enb$ modes, Appendix C gives perturbation matrix elements required to reproduce the numerical results of
\Sec{sec:Hom}.
Appendix D gives the evaluation of an important expression in Appendix A. Appendix E gives a new and valuable numerical recipe used in the production
of numerical results for this manuscript.

\section{The RSE Perturbation method}
\label{sec:AAAA1}

In this section I give a brief account of the derivation of the RSE \cite{More73,Bang78}. The method of derivation is built 
upon the Lippmann Schwinger Green's function equation. For a more detailed account of these derivations I direct the interested 
reader to Ref.\cite{More73,Bang78,ArmitagePRA14,DoostPRA13,DoostPRA14,DoostPRA12,MuljarovEPL10,Doost_Muljarov,Doost15,Doost16,Ge14}.

The spectral representation of the 2D system GF is modified to include this cut contribution as integral \cite{DoostPRA13}
\be \GFk(\br,\br')=\lefteqn{\sum_n}\int \,\frac{\En(\br)\otimes\En(\br')}{2 k(k-k_n)}\approx\sum_{\bar{n}}\frac{\phi_{\bar{n}}\Enb(\br)\otimes\Enb(\br')}{2 k(k-k_{\bar{n}})}
\label{ML3}\ee
In practice, the continuum of non-resonant states are discretised and included as cut poles in the perturbation basis. 
This problem is treated in depth in Appendix A, B and C. The combined index ${\bar{n}}$ is used to denote both real poles 
$k_n$ and cut poles $k_\alpha$ simultaneously. The weighting factors $\phi_{\bar{n}}$  are defined as follows
\be \phi_{\bar{n}}=\left\{
\begin{array}{ll}
\phi_n=1 & {\rm for\ real\ poles},\\
\phi_\alpha & {\rm for\ cut\ poles}\\
\end{array}
\right.
\ee
After modifying the expansion of the perturbed wave functions to include the cut poles to,
\be
 \Epn(\br) = \sum_{\bar{n}} \phi_{\bar{n}}b_{\bar{n}\nu} \Enb(\br) 
\ee
which is the solution to the equation,
\be \nabla\times\nabla\times\Epn(\br)=\varkappa^2_\nu\left[\heps(\br )+\Delta\heps(\br)\right]\Epn(\br)
\label{GFequ2} \ee
with outgoing BCs, I can repeat the derivation of Ref.\cite{DoostPRA13} and make use of the Lippmann Schwinger Green's function equation
\be \Epn(\br)=-\varkappa^2_\nu \int \GFk(\br,\br ')\Delta\heps(\br ')\Epn(\br ') d{\bf \br}'
\label{GFsol2} \ee
in conjuncture with the spectral GF
\be \GFk(\br,\br')=\lefteqn{\sum_n}\int \,\frac{\En(\br)\otimes\En(\br')}{2 k(k-k_n)}\approx\sum_{\bar{n}}\frac{\phi_{\bar{n}}\Enb(\br)\otimes\Enb(\br')}{2 k(k-k_{\bar{n}})}
\ee
to arrive at the modified the RSE Matrix equations,
\be \varkappa_\nu\sum _{\bar{n}'} \left(\delta_{\bar{n}\bar{n}'}+\sqrt{\phi_{\bar{n}}\phi_{\bar{n}}'}V_{\bar{n}{\bar{n}'}}/2\right)b_{\bar{n}'\nu}=k_n b_{\bar{n}\nu}
\label{RSE12D} 
\ee 
and
\be 
\sum_{\bar{n}'}\left(\frac{\delta_{\bar{n}{\bar{n}}'}}{k_{\bar{n}'}}+\frac{V_{\bar{n}{\bar{n}}'}}{2}\sqrt{\dfrac{\phi_{\bar{n}}\phi_{\bar{n}}'}{k_{\bar{n}} k_{\bar{n}'}}}\right)
c_{{\bar{n}'}\nu}=\frac{1}{\varkappa_\nu} c_{{\bar{n}}\nu}  
\label{RSE2D}
\ee
As always $V_{\bar{n}\bar{n}'}$ is given by
\be\label{Marks_VNM} 
V_{\bar{n}\bar{n}'}=\int \Enb(\br)\Delta\heps(\br)\Enpb(\br)\,d \br 
\ee

The perturbation method which I have briefly outlined was recently translated from Quantum Mechanics \cite{More73,Bang78} 
and has now become known as the Resonant-state-expansion.

\section{Maxwell's wave equation for magnetic field}
\label{sec:Maxwell2D}

In this section I discuss the numerical and analytic difficulties associated with constructing an electrodynamic RSE perturbation 
theory for magnetic resonances.

If we examine Maxwell's wave equation for magnetic field  $\mathbf{H(\br)}$,
\begin{multline}
- \nabla\times\nabla\times\mathbf{H(\br)}+{\epshat({\bf r})}^{-1}{\nabla\epshat({\bf r})}\times(\nabla\times\mathbf{H(\br)})
+k^2\epshat({\bf r})\mathbf{H(\br)}
\\
=\nabla\times\mathbf{J(\br)}-{\epshat({\bf r})}^{-1}{\nabla\epshat({\bf r})}\times\mathbf{J(\br)}\label{HJ}
\end{multline}
it would at first sight seem of no use to the RSE. In \Eq{HJ} we can see $\nabla\epshat({\bf r})$ 
which has prevented me from formulating an equation analogous to \Eq{GFsol2} relating the perturbed eigenmodes of magnetic field to their 
unperturbed GF and perturbation in the dielectric profile. Therefore as I am considering perturbations in the dielectric profile I am 
forced to formulate the RSE in terms of electric modes $\Enb(\br)$.

However even in light of the previous paragraph I do need to consider \Eq{HJ} in this paper because the cut poles 
cannot be normalised using \cite{DoostPRA14,Doost16}
\bea
\label{normaliz}
1+\delta_{k_n,0}&=&\int_A\En(\brd)\cdot\dfrac{\partial \omega^2\heps(\brd,\omega)}{\partial \omega^2}\bigg|_{\omega=\omega_n}\!\!\En(\brd)d{\brd}\\
&&+\lim_{k\rightarrow k_n}\oint_{L_A} \dfrac{\En\cdot\nabla\bE-\bE\cdot\nabla\En}{k^2-k^2_n} d{\bf L}\nonumber
\eea
The method of normalisation will thus make use of \Eq{HJ}. The reason I can't 
use \Eq{normaliz} for the cut poles is that they are not true resonances of the system. This problem can be intuitively understood if 
we assume we can normalise a single cut pole with \Eq{normaliz} and observe the resulting logical contradiction. If the discretisation 
of the cut is made finer by increasing the number of cut poles by several orders of magnitude, logically the normalisation of the chosen 
cut pole should be drastically reduced as it's weight is further shared between many of these extra poles. This final remark gives the 
contradiction because the normalisation calculated from \Eq{normaliz} cannot change, it is fixed.  

In light of the problems normalising cut poles I am forced to normalise the basis modes in this chapter by comparing spectral GFs of the 
form \cite{DoostPRA13} 
\be \GFk(\br,\br')=\lefteqn{\sum_n}\int \,\frac{\En(\br)\otimes\En(\br')}{2 k(k-k_n)}\approx\sum_{\bar{n}}\frac{\phi_{\bar{n}}\Enb(\br)\otimes\Enb(\br')}{2 k(k-k_{\bar{n}})}
\label{ML3}\ee
which will still hold true for cut poles, with the analytic GF of magnetic field calculated in Appendix B. 
Unlike the TM case where the analytic GF for the electric field $\GFEK(\br,\br')$ 
is available, I can only derive the analytic GF for the magnetic field $\GFHK(\br,\br')$ in the case of transverse electric (TE) polarisation. 
The reason for this is that a scalar GF equation is available for $\GFHK(\br,\br')$ but not for $\GFEK(\br,\br')$ in the TE case. Fortunately in 
Appendix A I have been able to derive a method of expressing the electric field basis modes $\Enb$ in terms of the magnetic field basis modes $\Hnb$. 

The Green's function for the magnetic component of an electrodynamic system is a tensor $\GFHK$ which 
satisfies the outgoing wave boundary conditions and Maxwell's wave equation with a delta function source term
\begin{multline}\label{eq:1}
- \nabla\times\nabla\times \GFHK(\br,\br')+ {\epshat({\bf r})}^{-1}{\nabla\epshat({\bf r})}\times(\nabla\times \GFHK(\br,\br'))
\\
+k^2\epshat({\bf r})\GFHK(\br,\br')
\\
=\hat{\mathbf{1}}\delta(\br-\br')
\end{multline}
following the derivation for the spectral representation of $\GFEK$ in Appendix B we find $\GFHK$ can be written as 
\be 
\GFHK(\br,\br')=\sum_n \frac{\phi_{\bar{n}}\Hnb(\br)\otimes\Hnb(\br')}{2k(k-k_{\bar{n}})}
\label{MK2M} 
\ee
where the field $\Hnb(\br)$ satisfies,
\begin{multline}
- \nabla\times\nabla\times\Hnb(\br)+ {\epshat({\bf r})}^{-1}{\nabla\epshat({\bf r})}\times(\nabla\times\Hnb(\br))
\\
+k^2\epshat({\bf r})\Hnb(\br)=0\label{RATNER}
\end{multline}
with outgoing BCs. In Appendix A I will make use of $\Hnb$ defined in \Eq{eq:1}, \Eq{MK2M} and \Eq{RATNER} to normalise $\Enb$.

I will briefly explain the idea behind the analytics in Appendix A and direct the interested reader to that relevant appendix where I give 
my detailed mathematics. The $\bE$ and $\bH$ fields are both generated by an electric current $\bJ$. By using an identical current to generate
$\bE$ and $\bH$ fields and letting the electric current amplitude tend to zero as its harmonic wave number of oscillation tends to $k_{\bar{n}}$
we can compare residues of $\GFHK$ and $\GFEK$ in order to relate normalised $\Enb$ and $\Hnb$. The $\Hnb$ modes, including the continuum modes, 
discussed in Appendix B, can be normalised by comparing the residue of the spectral GF and the residue of the analytic GF, derived in Appendix B.

\section{Unperturbed basis for 2D systems}
\label{sec:UPBF2D}

In this section I will use the approach outlined in \Sec{sec:Maxwell2D} to formulate the basis states for both TE, TM polarizations. 
For completeness of the basis I include longitudinal electric modes which are curl free static modes satisfying 
Maxwell's wave equation for $k_n=0$. Longitudinal electric modes have zero magnetic field.

Splitting off the time dependence $\propto e^{-i\omega t}$ of the electric fields ${\bf E}$ and ${\bf D}$ and magnetic field ${\bf H}$, the first pair of Maxwell's equations can be written in the form
\be
\nabla\times\bE=i k {\bf H}\,,\ \ \ \ \ \nabla\times{\bf H}=-i k {\bf D} - {\bf J}
\label{MEpair1}
\ee
where $k=\omega/c$ and ${\bf D}(\br)=\heps(\br){\bf E}(\br)$. Combining them leads to 
\be
- \nabla\times\nabla\times\mathbf{E(\br)}+k^2\heps(\br)\mathbf{E(\br)}=ik\mathbf{J(\br)}
 \label{EJequ0}
\ee
and
 \be
 \nabla\times\nabla\times\En(\br)=k_n^2\heps(\br)\En(\br)
 \label{me3DC2}
\ee which satisfy the outgoing wave boundary condition, and to 
\be \GFk(\br,\br')=\lefteqn{\sum_n}\int \,\frac{\En(\br)\otimes\En(\br')}{2 k(k-k_n)}\approx\sum_{\bar{n}}\frac{\phi_{\bar{n}}\Enb(\br)\otimes\Enb(\br')}{2 k(k-k_{\bar{n}})}
\label{ML3}\ee
for the corresponding GF. 

For $k_z=0$, these three groups of modes of a homogeneous dielectric cylinder can be written as
\bea\label{NTHERE}
{\rm TM:} & {\bf E}=f{{\bf e}_z} \,, \nonumber\\
{\rm TE:} & {\bf H}=f{{\bf e}_z}\,, \nonumber\\
{\rm LE:} & {\bf E}=-\nabla f\,, \nonumber
\eea
where $f(\br)$ is a scalar function satisfying the Helmholtz equation
\be
\nabla^2 f+k^2\varepsilon f=0
\label{Helm}
\ee
%The dielectric tensor has the form $\heps(\br)=\hat{\bf 1}\varepsilon(r)$, where
with the dielectric susceptibility of the sphere in vacuum
\be \varepsilon(\rho,\varphi)=\left\{
\begin{array}{lll}
n_r^2 & {\rm for} & \rho\leqslant R\,, \\
1 & {\rm for} & \rho>R
\end{array}
\right.\label{epsilon} \ee
and 
\be
f(\br)=R_m(\rho,k_n) \chi_m(\varphi)
\ee
The angular parts are defined by
\be
\chi_m(\varphi)=\left\{
\begin{array}{lll}
\pi^{-1/2}\sin(m\varphi) & {\rm if} & m<0\,, \\
(2\pi)^{-1/2} & {\rm if} & m=0\,, \\
\pi^{-1/2}\cos(m\varphi) & {\rm if} & m>0
\end{array}
\right. \label{chi-nC4} \ee
where there is no $m=0$ mode for the TE case. The $\chi_m(\varphi)$ are orthonormal according to
\be \int_0^{2\pi} \chi_m(\varphi)
\chi_{m'}(\varphi)d\varphi =\delta_{mm'} \ee

The radial components satisfy,
\be
\left[\dfrac{\partial^2}{\partial\rho^2}+\dfrac{1}{\rho}\dfrac{\partial}{\partial\rho}-\dfrac{m^2}{\rho^2}+\varepsilon(\rho)k^2\right]R_m(\rho,k)=0
\ee
and have the form
\be R_m(\rho,k)= \left\{
\begin{array}{lll}
J_m(n_r k\rho)/J_m(n_r kR) & {\rm for} & \rho\leqslant R\,, \\
H_m(k\rho)/H_m(kR) & {\rm for} & \rho>R \\
\end{array}
\label{R-analyt} \right. \ee
in which $J_m(z)$ and $H_m(z)\equiv H_m^{(1)}(z)$ are, respectively, the cylindrical Bessel and
Hankel functions of the first kind. $H_m(z)$ is a multiple valued function, with multiple values corresponding
to a range of possible boundary conditions at infinity. In order to obtain the correct boundary conditions for the 
RSE I provide a numerical recipe in Appendix E.

I treat in this work all polarisations, however only TM and LE polarisations mix due to the perturbations
treated being strictly scalar in dielectric permittivity. If the perturbation left 
the dielectric profile as a tensor then all polarisations would mix.  

The unperturbed RS wave functions factorise as
\bea\label{FINISHED_IT}
{\rm TM:} & \En(\br)=A^{\rm TM}R_m(\rho,k_n)\chi_m(\varphi){{\bf e}_z} \nonumber\\
{\rm TE:} & \Hn(\br)=A^{\rm TE}_m(k_n)R_m(\rho,k_n)\chi_m(\varphi){{\bf e}_z}
\eea
I normalized the wave functions from the analytic TM (TE) Green's function in Appendix \ref{App:GF} with normalisation constants $A(k)$
\be A^{\rm TM}=\frac{1}{R}\sqrt{\frac{2}{n_r^2-1}}
\label{A-normCTM}\ee
\be A^{\rm TE}_m=\sqrt{\frac{2k\left[J_m(n_rkR)\right]^2}{(n^2_r-1) \left[\dfrac{m^2}{k}\left[J_m(n_rkR)\right]^2+R^2k\left[J'_m(n_rkR)\right]^2\right]}}
\label{A-normCTE}\ee
The two boundary conditions at the surface of the cylinder, the continuity of the electric field
and its radial derivative, produce a secular equation for the RS wave number eigenvalues $k_n$,
which has the form
\be D^{\rm TM}_m(k_nR)=0\label{secular2} \ee
\be D^{\rm TE}_m(k_nR)=0\label{secular3} \ee
where
\be D^{\rm TM}_m(z)=n_rJ_m'(n_rz)H_m(z)-J_m(n_rz)H_m'(z) \label{Dfunc1} \ee
\be D^{\rm TE}_m(z)=J_m'(n_rz)H_m(z)-n_rJ_m(n_rz)H_m'(z) \label{Dfunc2} \ee
and $J_m'(z)$ and $H_m'(z)$ are the derivatives of $J_m(z)$ and $H_m(z)$, respectively. Here $z$ represents a complex argument, as opposed to the spatial coordinate used earlier.

In cylindrical coordinates, the electric vector field $\bE(\br)$ can be written as
\be \bE(\rho,\varphi,z)=E_{\rho}{\bf e}_r+E_{\varphi}{\bf e}_{\varphi}+E_{z}{\bf e}_{z}= \begin{pmatrix}
E_{\rho} \\
E_{\varphi}\\
E_{z}\\
\end{pmatrix} \label{E-factC} \ee
Therefore the TM modes can be written as
\be \mathbf{E}^{\rm TM}_{\bar{n}}(\br)=A^{\rm TM}\left(
\begin{array}{ccc}
0\\[3pt]
0\\[7pt]
R_m(\rho,k_{\bar{n}})\chi_m(\varphi)\\
\end{array}
\right)\label{ETMFORM}\ee

Since I am considering a homogeneous micro-cylinder and TE modes, I can now calculate the normalised $\Enb$ from the $\Hnb$ using the following 
simple relations derived in Appendix A
\be
\nabla\times\Hnb(\br)={M_{k_{\bar{n}}}}{k^2_{\bar{n}}}\Enb(\br)
\ee
which gives, inside the homogeneous cylinder only,
\be \mathbf{E}^{\rm TE}_{\bar{n}}(\br)=\dfrac{A^{\rm TE}}{M_{k_{\bar{n}}}k^2_{\bar{n}}}\left(
\begin{array}{ccc}
\dfrac{R_m(\rho,k_{\bar{n}})}{\rho}\dfrac{d\chi_m(\varphi)}{d\varphi}\\[3pt]
-\dfrac{\partial R_m(\rho,k_{\bar{n}})}{\partial \rho}\chi_m(\varphi)\\[7pt]
0\\
\end{array}
\right)\label{ETMFORM}\ee
The unknown $M_{k_{\bar{n}}}$ is calculated analytically in Appendix A and Appendix D.

The $k=0$ LE modes are given by, inside the homogeneous cylinder only,
\be
\mathbf{E}^{\rm LE}_{\bar{n}}(\br)=\sqrt{|m|(n^2_r-1)}\lim_{k_n\rightarrow 0}\mathbf{E}^{\rm TE}_{\bar{n}}(\br)
\ee
which follows from \Eq{normaliz}, in the limit of $k=0$, noting the fast attenuation of the RS far field in this $k=0$ case.

For both the TM and TE case, treated here, the full GF of the homogeneous dielectric cylinder, 
which is defined via Maxwell's equation with a line current source term, is given by $\GFk=G_k\,{{\bf e}_z}\otimes {{\bf e}_z}$,
in which
\be G_k(\brho,\brho')=\sum_m G_m(\rho,\rho';k) \chi_m(\varphi)\chi_m(\varphi')
\label{GF-TM} \ee
and the radial components have the following spectral representation
\bea
G_m(\rho,\rho')&=&\sum_n\frac{\left[ A_m(k_n)\right]^2R_m(\rho,k_n)R_m(\rho',k_n)}{2k(k-k_n)}\nonumber\\
&&+\int_{-i\infty}^0 \frac{\left[ A_m(k)\right]^2R_m(\rho,k')R_m(\rho',k')}{2k(k-k')}\sigma_m(k') dk'\nonumber\\
&\equiv&\lefteqn{\sum_n} \int \,\frac{\left[ A_m(k_n)\right]^2R_m(\rho,k_n)R_m(\rho',k_n)}{2k(k-k_n)} \label{Gm} \eea
derived in Appendix~\ref{App:GF}. For TE modes I take $A_m(k)=A^{\rm TE}_m(k)$ and for the TM modes I take $A_m(k)=A^{TM}_m$, then
\be
\sigma_m(k)=\frac{8J_m(n_rkR)^2}{\pi^2 kR^2 D_m^+(kR)D_m^-(kR)}\frac{1}{A_m(k)^2} 
\label{Sigma}\ee
Here $D^{\pm}_m(z)$ are the two limiting values of $D_m(z')$ for
$z'$ approaching point $z$ on the cut from its different sides Re$\,z'\gtrless 0$. $D^{\pm}_m(z)$ are taken to be $D^{\rm TE}_m$ or $D^{\rm TM}_m$
in the case of TE or TM cut pole respectively.

\begin{figure}[t]
\includegraphics*[width=0.95\columnwidth]{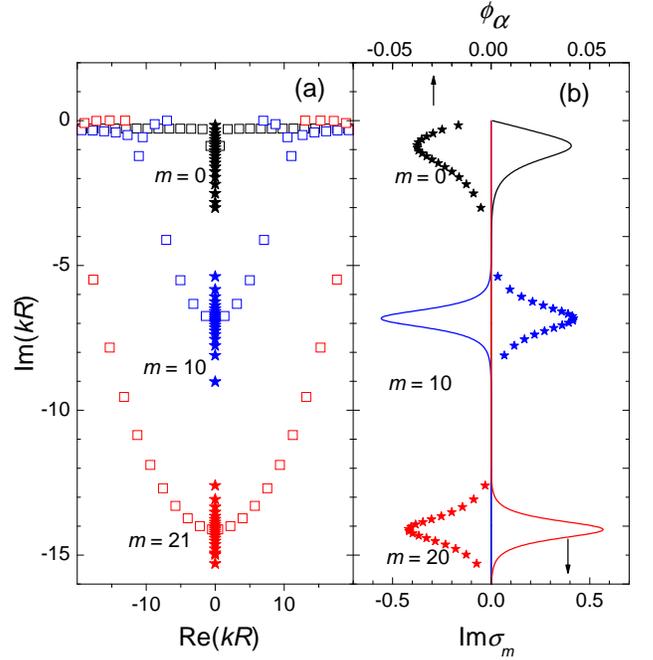}
\caption{(a): Cut poles $k_\alpha$ (stars)  representing the cut of the GF of a homogeneous dielectric cylinder
with ${n_r}=2$, in the complex wave-number plane for $m=0$, 11, and 20. Normal poles
$k_n$ (open squares) are also shown. (b): Cut pole density $\sigma_m(k)$ (solid curves) and the cut pole strength $\phi_\alpha^{(m)}$ (stars), for the same values of $m$.
}\label{fig:Fict}
\end{figure}

I use the same method to discretised this continuum into cut poles $k_\alpha$ and cut pole strengths $\phi_\alpha^{(m)}$ as in my Ref.\cite{DoostPRA12}
\bea k_\alpha^{(m)}=\left.\int_{q_\alpha^{(m)}}^{q_{\alpha+1}^{(m)}}k\sigma_m(k)dk\right/\phi_\alpha^{(m)}
\label{k-alpha} \eea
where the cut pole strength $\phi_\alpha^{(m)}$ is defined as
\be
\phi_\alpha^{(m)}=\int_{q_\alpha^{(m)}}^{q_{\alpha+1}^{(m)}}\sigma_m(k)dk
\label{phi-m-alpha}
\ee

An example of cut poles assigned for $m=0$, 10, and 20 is given in \Fig{fig:Fict}(a).
The cut poles contribute to the RSE in the same way as the normal poles.

In the numerical calculation in Sec.\,\ref{sec:Hom}, I use $N_c^{(m)}\sim N^{(m)}$, where $N^{(m)}$ is the number of normal poles in the basis for the 
given $m$ and $N_c^{(m)}$ is the number of cut poles, in order to demonstrate the convergence towards the
exact solution.

The next section discusses results of the RSE for different effective 2D systems. We
consider a homogeneous dielectric cylinder of radius $R$ and refractive index $n_r=2$
($\varepsilon=4$) with a homogeneous perturbation of the
whole cylinder in \Sec{sec:Hom}. Explicit forms of the
matrix elements for these perturbations and details of their calculation are given in
Appendix C.

\begin{figure}[b]
\includegraphics*[width=0.95\columnwidth]{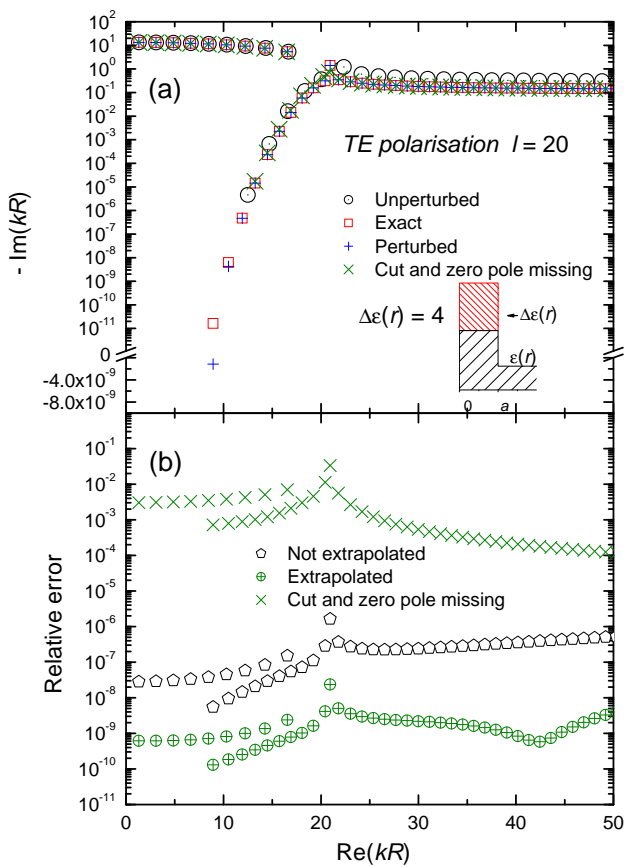}
\caption{(a): Perturbed RS wave numbers for the homogeneous perturbation Eq.\,(\ref{eps-hom})
calculated via the RSE with $N=800$ (only sine modes are shown). The perturbed poles with (+) and without ($\times$) the cut
contribution are compared with the exact solution (open squares). Unperturbed wave numbers are
also show (open circles with dots). Inset: Dielectric constant profile for the unperturbed and perturbed
systems. (b): Relative error in the calculated
perturbed wave numbers with (heptagons) and without (triangles) contribution of the cut. Relative
error for a simulation including the cut and improved by extrapolation is also shown (crossed
circles). }\label{fig:F2}
\end{figure}

\section{Homogeneous Cylinder Perturbation}
\label{sec:Hom} 

In this section I present my numerical results which support the validity of the analytic results relating to the 
$2$-dimensional TE RSE which I have provided in this manuscript.

The perturbation we consider in this section is a homogeneous change of
$\varepsilon$ over the whole cylinder, given by
\be \Delta\varepsilon(\rho,\varphi)=\Delta\varepsilon \theta(R-\rho)=\left\{
\begin{array}{cl}
\Delta\varepsilon & \text{for\ \  } \rho \leqslant R\,,\\
0 &\text{for\ \  } \rho > R
\end{array} \right.
\label{eps-hom} \ee
with the strength $\Delta\varepsilon=4$ used in the numerical calculation. For
$\varphi$-independent perturbations, modes with different azimuthal number $m$ are decoupled, and so
are even and odd (cosine and sine). I show only the sine modes here, and use
for illustration $m=20$. The matrix elements of the perturbation are calculated
analytically and given by \Eq{Marks_VNM}. The homogeneous
perturbation does not change the symmetry of the system, so that the perturbed modes obey the same
secular equation \Eq{secular3} with the refractive index $n_r$ of the cylinder changed to
$\sqrt{n_r^2+\Delta\varepsilon}$, and thus the perturbed wave numbers $\varkappa_\nu$ calculated
using the RSE can be compared with the exact values $\varkappa^{\rm (exact)}_\nu$.

We choose the basis of RSs for the RSE in such a way that for the given azimuthal number $m$ and the given number of normal RSs $N$ we find all normal poles $|k_n|<k_{\rm max}(N)$ with a suitably chosen maximum wave vector $k_{\rm max}(N)$ and then add the cut poles.
We find that as we increase $N$, the relative error $\bigl|{\varkappa_\nu}/{\varkappa^{\rm
(exact)}_\nu}-1\bigr|$ decreases as $N^{-3}$. Following the procedure described in
Ref.\cite{DoostPRA12} we can extrapolate the perturbed wave numbers. The resulting
perturbed wave numbers are shown in \Fig{fig:F2}. The perturbation is strong, creating 3 additional
WGMs with $m=20$ having up to 4 orders of magnitude narrower linewidths. For $N=800$, the RSE reproduces about 100 modes to a relative error in the $10^{-7}$ range, which is decreasing by one or two orders of magnitude after extrapolation. The contribution of the cut and zero frequency mode is
significant: Ignoring these modes leads to a relative error of the poles in the $10^{-3}$ range. The fact that the relative error improves by 4-5 orders of magnitude after taking into account the cut in the form of the cut poles
shows the validity of the reported analytical treatment of cuts in the RSE, and the
high accuracy of the discretisation method into cut poles.

\section{Summary}\label{sec:summary}
I have applied the resonant state expansion (RSE) to effective two-dimensional (2D) open optical
systems, such as dielectric micro-cylinders and micro-disks with perturbations, by finding a formula linking the normalisation of resonant states appearing in the magnetic and electric Green's functions which describe identical systems. These analytic formulas allow us to take advantage of an analytic Green's function of the magnetic field of the homogeneous micro-cylinder to normalise the modes appearing in the spectral representation of the Green's function for the electric field, electric modes which when normalised form the basis of the RSE. 

Using the analytically known basis of resonant states (RSs) of an ideal homogeneous dielectric
cylinder -- a complete set of eigenmodes satisfying outgoing wave boundary conditions -- I have
treated an effectively 1D homogeneous perturbations. I investigated the convergence for this
perturbations and compared the RSE with analytic solutions. 
I have found agreement between the RSE and analytic solutions.

\acknowledgments 

The work in this paper is completely my own original mathematical derivations and numerical data. This work was completed in 2013, and shortly after its 
completion I presented a draft of this manuscript to my then collaborator Dr E. A. Muljarov. Upon Dr E. A. Muljarov's insistence (as my PhD supervisor and 
manager), a reformulated version of the mathematics 
in Appendix A was generated in collaboration with Dr E. A. Muljarov and presented in my PhD Thesis. I do not make use of any part of these reformulations of my 
original work in this manuscript, hence I am the sole author of this work. 
\appendix
\section{Normalised electric modes expressed in terms of normalised magnetic modes}
\label{App:ML}

In this appendix I provide the analytic formula relating the normalised $\Enb$ and normalised $\Hnb$ modes. The basic principals of my derivation are
as follows, let us consider the case where the $\bE$ and $\bH$ fields are both generated by a oscillating 
electric current $\bJ$. By using an identical current to generate
$\bE$ and $\bH$ fields and letting the current tend to zero as its harmonic wavenumber of oscillation tends to the resonant wavenumber $k_{\bar{n}}$
we can compare residues of $\GFHK$ and $\GFEK$ in order to relate normalised $\Enb$ and $\Hnb$. The $\Hnb$ modes, including the continuum modes, 
discussed in Appendix B, can be normalised by comparing the residue of the spectral GF and the residue of the analytic GF, derived in Appendix B. The 
detailed mathematics will now be thoroughly explained.

Our starting point for finding a relation between the correct normalisation of the modes appearing in the 
spectral Green's function for the magnetic $\mathbf{H(\br)}$-field and electric $\mathbf{E(\br)}$-field of a homogeneous dielectric micro-cylinder 
is Maxwell's equations of electrodynamics,
\be
\nabla\times\bE=i k {\bf H}\,,\ \ \ \ \ \nabla\times{\bf H}=-i k {\bf D} - {\bf J}
\label{MARW}
\ee
where we have assumed harmonic oscillation of the driving current with respect to time, so that the time derivative in \Eq{MARW} 
can be replaced with $ik$.

For the $\mathbf{E(\br)}$-field with a current source $\mathbf{J(\br)}$ oscillating at frequency $k$ we have according to Maxwell, 
 \be
- \nabla\times\nabla\times\mathbf{E(\br)}+k^2\varepsilon\mathbf{E(\br)}=ik\mathbf{J(\br)}
 \label{EJequ}
\ee
and also the corresponding wave equation with a source for the $\mathbf{H(\br)}$-field,
\begin{multline}\label{HJequ}
- \nabla\times\nabla\times\mathbf{H(\br)}+ \dfrac{\nabla\varepsilon(\br)}{\varepsilon(\br)}\times(\nabla\times\mathbf{H(\br)})
+k^2\varepsilon(r)\mathbf{H(\br)}
\\
=\nabla\times\mathbf{J(\br)}-\dfrac{\nabla\varepsilon(\br)}{\varepsilon(\br)}\times\mathbf{J(\br)}
\end{multline}
The GF for \Eq{EJequ} is,  
\be
- \nabla\times\nabla\times \GFEk(\br,\br')+k^2\hat{\boldsymbol{\varepsilon}}({\bf r})\GFEk(\br,\br')=\hat{\mathbf{1}}\delta(\br-\br')
 \label{GFEequ}
\ee
and for \Eq{HJequ} 
\begin{multline}\label{APP_APP}
- \nabla\times\nabla\times \GFHK(\br,\br')+ {\epshat({\bf r})}^{-1}{\nabla\epshat({\bf r})}\times(\nabla\times \GFHK(\br,\br'))
\\
+k^2\epshat({\bf r})\GFHK(\br,\br')
\\
=\hat{\mathbf{1}}\delta(\br-\br')
\end{multline}
The formulas for the $\mathbf{E(\br)}$-field  and  $\mathbf{H(\br)}$-field in terms of their respective Green's functions are given as the
following convolutions with the source,
\be 
\mathbf{E(\br)}=\int\GFEk(\br,\br')i k\mathbf{J(\br)} d\br'\label{GEA1equ}
\ee

\be 
\mathbf{H(\br)}=\int\GFHk(\br,\br')\left[ \nabla\times\mathbf{J(\br')}-\dfrac{\nabla\varepsilon(\br')}{\varepsilon(\br')}\times\mathbf{J(\br')}  \right] d\br'\label{GHA1equ}
\ee

Since at this stage $\mathbf{J(\br)}$ is arbitrary we can set $\mathbf{J}(\br)=0$ when $\rho>R_{-}$ where $R_{-}$ is some radius $R_{-}<R$. 
$R$ is the radius of the micro-cylinder.
Also since we choose our basis system to be a homogeneous micro-cylinder, importantly for \Eq{EEPND} $\nabla\varepsilon(\rho,\varphi)=0$ when $\rho<R$. 
At this stage $k$ is also arbitrary so let it be $k=k_{\bar{n}}+\Delta k$, $\Delta k\rightarrow 0$. 
To simplify our discussion let $\mathbf{W(r)}$ be a function  related to  $\mathbf{J(\br)}$ by,  
\be\label{DONE_GOOD}
ik\mathbf{J}(\br)=2k_{\bar{n}}\Delta k \mathbf{W(\br)}
\ee 
\Eq{DONE_GOOD} allows $\mathbf{J(\br)}$ to meet the requirements that I have just set out.

In light of \Eq{DONE_GOOD} \Eq{GEA1equ} becomes
\begin{multline}\label{HEPND}
\lim_{\Delta k\rightarrow0}\int_{R_{-}}\left[\sum_{\bar{m}} \frac{\phi_{\bar{m}}\Emb(\br)\otimes\Emb(\br')}{2k_{\bar{m}}(k_{\bar{n}}-k_{\bar{m}}+\Delta k)}\right]2k_{\bar{n}}\Delta k \mathbf{W(r')} d\br'
\\
=\phi_{\bar{n}}\mathbf{E_{\bar{n}}(\br)}\int\Enb(\br')\cdot \mathbf{W(r')} d\br'=\mathbf{E(r)}
\end{multline}
where we have replaced $\GFEk(\br,\br')$ with its spectral representation 

\be \GFk(\br,\br')=\lefteqn{\sum_n}\int \,\frac{\En(\br)\otimes\En(\br')}{2 k(k-k_n)}\approx\sum_{\bar{n}}\frac{\phi_{\bar{n}}\Enb(\br)\otimes\Enb(\br')}{2 k(k-k_{\bar{n}})}
\label{ML3A}\ee

Similarly \Eq{GHA1equ} becomes
\begin{multline}\label{EEPND} 
\mathbf{H(r)}=\lim_{\Delta k\rightarrow0}\int_{R_{-}}\left[\sum_{\bar{m}} \frac{\phi_{\bar{m}}\Hmb(\br)\otimes\Hmb(\br')}{2k_{\bar{m}}(k_{\bar{n}}-k_{\bar{m}}+\Delta k)}\right]
\\
\cdot\left[\dfrac{2k_{\bar{n}}\Delta k\nabla\times\mathbf{W(r')}}{ik}\right]d\br'
\\
=\phi_{\bar{n}}\mathbf{H_{\bar{n}}(r)}\int_{R_{-}} \dfrac{\Hnb(\br')\cdot(\nabla\times\mathbf{W(r')})}{ik_{\bar{n}}} d\br'
\end{multline}
where we have replaced $\GFHk(\br,\br')$ with its spectral representation 
\be \GFHK(\br,\br')=\sum_n \frac{\phi_{\bar{n}}\Hnb(\br)\otimes\Hnb(\br')}{2k(k-k_{\bar{n}})}
\label{MK2} \ee

\Eq{EEPND} and \Eq{HEPND} can be substituted in \Eq{MARW} to give,
\begin{multline}\label{LINK}
\left[\nabla\times\mathbf{\Enb(\br)}\right]\int_{R_{-}}\Enb(\br')\cdot \mathbf{W(r')} d\br'
\\
=\mathbf{\Hnb(r)}\int_{R_{-}} \Hnb(\br')\cdot(\nabla\times\mathbf{W(r')}) d\br'
\end{multline}
We now make the substitution $\mathbf{W(\br)}=\Gamma(\rho)\mathbf{\Enb(\br)}$, where,
\be \Gamma(\rho)=\left\{
\begin{array}{cl}
1 & \text{for\ \  } \rho \leqslant R_{-}\,,\\
0 &\text{for\ \  } \rho > R_{-}
\end{array} \right.
\ee
We can calculate the curl of $\mathbf{W(\br)}$ analytically. First note for a given mode we can evaluate \Eq{MARW} to be equivalently (in this case)
\be
\nabla\times\Enb(\br)=\dfrac{n^2_r\Hnb(\br)}{M_{k_{\bar{n}}}}
\label{RMK}
\ee 
where ${M_{k_{\bar{n}}}}$ is an unknown to be found.

Using \Eq{RMK} we can write the following identity for $\mathbf{W(\br)}$: 
\begin{multline}\label{WWW} 
\nabla\times\mathbf{W(\br)}=\nabla\Gamma(\rho)\times\Enb(\br)
+\Gamma(\rho)\nabla\times\Enb(\br)
\\
=-\delta(\rho-R_{-})e_{\rho}\times\mathbf{\Enb(\br)}+\Gamma(\rho)\dfrac{n^2_r\Hnb(\br)}{M_{k_{\bar{n}}}}
\end{multline}

Substituting \Eq{WWW} into \Eq{LINK} and cancelling common factors of $\Hnb(\br)$ gives
\begin{multline}\label{LINKS} 
\left[\dfrac{n^2_r}{M_{k_{\bar{n}}}}\right]\int_{R_{-}}\Enb(\br')^2 d\br'
\\
=-\int_{R_{-}} \Hnb(\br')\cdot\delta(\rho'-R_{-})e_{\rho}\times\mathbf{\Enb(\br')} d\br'+
\\+\int_{R_{-}}\Gamma(\rho')\dfrac{n^2_r\left[\Hnb(\br')\right]^2}{M_{k_{\bar{n}}}} d\br'
\end{multline}

Substituting \Eq{ETMFORM} into \Eq{LINKS} 
\begin{multline}\label{LINKS2} 
\int_{R_{-}}\Enb(\br')^2 d\br'\\
=\int^{\pi}_{-\pi}R_{-}d\varphi \chi_m(\varphi) \chi_m(\varphi)\cdot\dfrac{A^2_{\bar{n}}J_m(nk_{\bar{n}}R_{-})}{k^2_{\bar{n}}n^2_r}\dfrac{\partial{J_m(nk_{\bar{n}}R_{-})}}{\partial{\rho}}
\\
+\int_{R_{-}}\Hnb(\br')^2 d\br'
\end{multline}
\Eq{ETMFORM} was generated by \Eq{BANG_BANG}, where $M_{k_{\bar{n}}}$ is the unknown we are trying to find.
In \Eq{LINKS2} we took
\be
\Hnb(\br)=A_{\bar{n}}J_m(nk_{\bar{n}}\rho)\chi_m(\varphi){{\bf e}_z}
\ee
and re-calculated $\Enb$ according to \Eq{BANG_BANG}.

Calculating the first integral in \Eq{LINKS2} by making substitutions from  \Eq{ETMFORM}   
\begin{multline}\label{STUFF21}
\int_{R_{-}}\Enb(\br')^2 d\br'\\
=\dfrac{A(k_{\bar{n}})^2}{{M^2_{k_{\bar{n}}}}k^4_{\bar{n}}}\int_{R_{-}}\left[\dfrac{m^2J^2_m(nk_{\bar{n}}\rho')}{\rho'}+\rho'\left[\dfrac{\partial{J_m(nk_{\bar{n}}\rho')}}{\partial{\rho}}\right]^2\right]d\rho'
\end{multline}
having made use of
\be\label{BANG_BANG}
\nabla\times\Hnb(\br)=M_{k_{\bar{n}}}k^2_{\bar{n}}\Enb(\br)
\ee
which comes from taking the curl of \Eq{RMK} and using the definition of RSs, specifically, inside the homogenous cylinder, 
\be
\nabla\times\nabla\times\Enb(\br)=k^2_{\bar{n}}n^2_r\Enb(\br)
\ee

\Eq{STUFF21} is not instantly recognisable as a standard integral, however if we make use of the two following general properties of Bessel function, 
\be
J_{m-1}(z)-J_{m+1}(z)=2{J}_m'(z)
\ee
\be
J_{m-1}(z)+J_{m+1}(z)=\dfrac{2m}{z}{J}_m(z)
\ee
we see that
\begin{multline}\label{TE1}
\int_{R_{-}}\Enb(\br')^2 d\br'\\
=\dfrac{A(k_{\bar{n}})^2n^2_r}{2M^2_{k_{\bar{n}}}k^2_{\bar{n}}}\int_{R_{-}}\left[J^2_{m-1}(n_rk_{\bar{n}}\rho')+J^2_{m+1}(n_rk_{\bar{n}}\rho')\right]\rho'd\rho'
\end{multline}
which is a standard integral given in Appendix C.

In the last integral in \Eq{LINKS2} we can make substitutions from \Eq{NTHERE} and \Eq{FINISHED_IT} to obtain the following standard integral given in Appendix C,  
\be
\int_{R_{-}}\Hnb(\br')^2 d\br'=A(k_{\bar{n}})^2\int_{R_{-}}J^2_{m}(nk_{\bar{n}}\rho')\rho'd\rho'
\label{TE2}
\ee

Finally we use \Eq{TE1} and \Eq{TE2} to solve \Eq{LINKS2} for $M_{k_{\bar{n}}}$
\begin{multline}\label{HELIOS}
\dfrac{2{M^2_{k_{\bar{n}}}}}{n^4_r}=
\\
=\dfrac{\int\left[J^2_{m-1}(n_rk_{\bar{n}}\rho')+J^2_{m+1}(n_rk_{\bar{n}}\rho')\right]\rho'd\rho'}{R_{-}J_m(n_rk_{\bar{n}}R_{-})\dfrac{\partial{J_m(n_rk_{\bar{n}}R_{-})}}{\partial{r}}+k^2_{\bar{n}}\varepsilon\int J^2_{m}(nk_{\bar{n}}\rho')\rho'd\rho'}
\\
=\dfrac{2}{n^2_rk^2_{\bar{n}}}
\end{multline}
or
\be
M^2_{k_{\bar{n}}}=\dfrac{n^2_r}{k^2_{\bar{n}}}
\ee
I evaluate analytically the integral in \Eq{HELIOS} in Appendix D. 

Hence our analytics in this section have reached their conclusion. We can now mathematically relate normalised $\Enb$ modes to normalised $\Hnb$ modes
with the analytic formula,
\be\label{LENNON}
\nabla\times\Hnb(\br)=M_{k_{\bar{n}}}k^2_{\bar{n}}\Enb(\br)
\ee

I make use of this valuable derived \Eq{LENNON}
in Appendix B, \Sec{sec:Maxwell2D}, \Sec{sec:UPBF2D} and \Sec{sec:Hom}, 
to translate the correctly normalised $\Hnb$ modes therein derived, into correctly 
normalised $\Enb$ modes.

\section{Green's function of a homogeneous cylinder}
\label{App:GF}

In this appendix I nearly exactly repeat the derivation I made with E. A. Muljarov in Ref.\cite{DoostPRA13} and derive an
analytic GF for $2$-dimensional cylindrical homogeneous electromagnetic resonators. 

The only difference between the analytics here
and the analytics of Appendix B Ref.\cite{DoostPRA13} is that here I am restricted to the magnetic $\bH$-field GF of the $2$-dimensional TE modes of the 
cylinder, while in Appendix B of Ref.\cite{DoostPRA13} we were restricted to the $\bE$-field TM modes of a mathematically identical system. 

The reason 
for the difference in approach between this appendix and Ref.\cite{DoostPRA13}
is that to formulate the $\bE$-field TE modes into effectively $1$-dimensional equations is currently an unsolved problem. To generate the analytic
GF of the electrodynamic problem, using the methods in this appendix, requires the recasting of the equation into a set of decoupled $1$-dimensional problems.

The TE component of the GF of a homogeneous cylinder in vacuum satisfies the following equation for magnetic field
\begin{multline}
- \nabla\times\nabla\times \GFHk(\br,\br')+ \dfrac{\nabla\varepsilon({\bf r})}{\varepsilon({\bf r})}\times(\nabla\times \GFHk(\br,\br'))
\\
+k^2\varepsilon({\bf r})\GFHk(\br,\br')=\hat{\mathbf{1}}\delta(\br-\br')
\end{multline}
with
\be \varepsilon(\rho)=\left\{
\begin{array}{lll}
n_r^2 & {\rm for}&\rho \leqslant R\,,\\
1& {\rm for} &\rho >R
\end{array}
\right. \ee
Using the angular basis \Eq{chi-nC4} the GF can be written as
\be G^H_k(\brho,\brho')=\frac{1}{\sqrt{\rho\rho'}}\sum_m \tilde{G}^H_m(\rho,\rho';k)
\chi_m(\varphi)\chi_m(\varphi') \ee similar to \Eq{GF-TM}. Note that we redefined here the radial part as
$\tilde{G}^H_m(\rho,\rho';k)=\sqrt{\rho\rho'}{G}^H_m(\rho,\rho';k)$ which satisfies
\begin{multline}
\left[\frac{d^2}{d\rho^2}-\frac{m^2-1/4}{\rho^2}+\dfrac{1}{\varepsilon(\rho)}\dfrac{d\varepsilon(\rho)}{d\rho}\dfrac{1}{2\rho}\right]\tilde{G^H}_m(\rho,\rho';k)+
\\
+\left[k^2\varepsilon(\rho)-\dfrac{1}{\varepsilon(\rho)}\dfrac{d\varepsilon(\rho)}{d\rho}\frac{d}{d\rho}\right]\tilde{G^H}_m(\rho,\rho';k)=\delta(\rho-\rho')
\end{multline}
Using two linearly independent solutions $f_m(\rho)$ and $g_m(\rho)$ of the corresponding
homogeneous equation which satisfy the asymptotic boundary conditions
$$
\begin{array}{lll}
f_m(\rho)\propto \rho^{m+1/2}&{\rm for}&\rho\to 0\,,\\
g_m(\rho)\propto e^{ik\rho}& {\rm for}&\rho\to \infty
\end{array}
$$
the GF can be expressed as
\be\label{WRON2} \tilde{G^H}_m(\rho,\rho';k)=\frac{f_m(\rho_<)g_m(\rho_>)}{W(f_m,g_m)} \ee
in which $\rho_<=\min\{\rho,\rho'\}$, $\rho_>=\max\{\rho,\rho'\}$, and the Wronskian $W(f,g)= f
g'-f'g$\,. For TE polarization, a suitable pair of solutions is given by
\bea f_m(\rho)&=&\sqrt{\rho}\cdot\left\{
\begin{array}{ll}
J_m(n_r\rho k)\,, & \rho \leqslant R\,,\\
a_m J_m(\rho k) +b_m H_m(\rho k)\,,& \rho >R
\end{array}
\right.
\nonumber \\
g_m(\rho)&=&\sqrt{\rho}\cdot\left\{
\begin{array}{ll}
c_m J_m(n_r\rho k) +d_m H_m(n_r\rho k)\,,& \rho \leqslant R\,,\\
H_m(\rho k)\,,& \rho >R
\end{array}
\right. \nonumber \eea
where
\bea a_m(k)&=&\bigl[H_m(x)J_m'(n_rx)-n_rH_m'(x)J_m(n_r x)\bigr]\pi ix/2n_r\,,
\nonumber \\
b_m(k)&=&\bigl[n_rJ_m'(x)J_m(n_rx)-J_m(x)J_m'(n_r x)\bigr]\pi ix/2n_r\,,
\nonumber \\
c_m(k)&=&\bigl[n_rH_m'(x)H_m(n_rx)-H_m(x)H_m'(n_r x)\bigr]\pi in_rx/2\,,
\nonumber \\
d_m(k)&=&\bigl[H_m(x)J_m'(n_rx)-n_r H_m'(x)J_m(n_r x)\bigr]\pi in_rx/2 \nonumber \eea
with $x=kR$. The Wronskian is calculated to be
\be\label{Wron}
W(f_m,g_m)= 2in^2_ra_m(k)=-nxD_m(x) /\pi,
\ee
with $D_m(x)$ defined in \Eq{Dfunc2}. Inside the cylinder, the GF then takes the form
\bea
\hspace{-1cm}\tilde{G^H}_m(\rho,\rho';k)&=&\frac{\pi}{2i} \sqrt{\rho\rho'} \biggl[J_m(n_r
k\rho_<)H_m(n_r k\rho_>)
\nonumber\\
&&\left.-\frac{c_m(k)}{a_m(k)}J_m(n_r k\rho_<)J_m(n_r k\rho_>)\right]
\label{GF-Hankel}
\eea

The GF has simple poles $k_n$ in the complex $k$-plane which are the wave vectors of RSs, given by
$a_m(k_n)=0$, an equation equivalent to Eq.\,(\ref{secular3}). The residues
 ${\rm Res}_n$ of the GF at these poles are calculated using
\begin{multline}
 r_m(k_n)=\left.-\frac{c_m(k)}{\frac{d}{dk} a_m(k)}\right|_{k=k_n}
\\
=\frac{2i}{\pi(n^2_r-1) \left[\dfrac{m^2}{k_n}\left[J_m(nk_nR)\right]^2+R^2k_n\left[J'_m(nk_nR)\right]^2\right]} \label{Res} \end{multline}
which can be seen by noting that inside the resonantor \Eq{WRON2} and \Eq{MK2} are equal as $k\rightarrow k_n$ as they are equal for all $k$,
at every point in space inside the resonator.

In addition to the poles, the GF has a cut in the complex $k$-plane along the negative imaginary half-axis.
The cut is due to the Hankel function $H_m(z)$. 

Owing to the cut of the Hankel function $H_m(z)$  the GF also
has a cut along the negative imaginary half-axis in the complex $k$-plane, so that on both sides
of the cut $\tilde{G}_m$ takes different values: $\tilde{G}^+_m$ one the right-hand side and
$\tilde{G}^-_m$ one the left-hand side of the cut. The step
$\Delta\tilde{G}_m=\tilde{G}^+_m-\tilde{G}^-_m$ over the cut can be calculated using the
corresponding difference in the Hankel function:
$$
\Delta H_m(z)=H^+_m(z)-H^-_m(z)=4J_m(z)
$$
The result is
\be \Delta\tilde{G^H}_m(\rho,\rho';k)=\frac{\pi}{2i}\sqrt{\rho\rho'}J_m(n_r
k\rho_<)J_m(n_r k\rho_>)\Delta Q_m(k) \label{deltaG} \ee
where
\be \Delta
Q_m(k)=\left[4-\frac{c_m^+}{a_m^+}+\frac{c_m^-}{a_m^-}\right]=-\left(\frac{4}{\pi
kR}\right)^2\frac{1}{D_m^+(k)D_m^-(k)} \ee
with $D_m(k)$ given by \Eq{Dfunc2}.

\bea
&&\oint\frac{\tilde{G^H}_m(\rho,\rho';k')}{k-k'} d k'= \int^{-i\infty}_0\frac{\tilde{G^H}^+_m d k'}{k-k'}+\int_{-i\infty}^0\frac{\tilde{G^H}^-_m d k'}{k-k'}\nonumber\\
&&=2\pi i \sum_n \frac{{\rm Res}_n}{k-k_n}-2\pi i  \tilde{G^H}_m(\rho,\rho';k) \label{integrals}
\eea
Note that in second part of the above equation we have made use of the residue theorem, expressing
the closed-loop integral in the left-hand side in terms of a sum over residues at all poles inside
the contour. Using \Eq{integrals} the GF can be expressed as
\be
\tilde{G}^H_m(\rho,\rho';k)=\sum_n \frac{{\rm Res}_n}{k-k_n}+\frac{1}{2\pi i}
\int_{-i\infty}^0\frac{\Delta\tilde{G}^H_m(\rho,\rho';k') d k'}{k-k'} \label{GF-exp} \ee
which is
a generalization of the Mittag-Leffler theorem. 

In conclusion then, the residues ${\rm Res}_n$ of the GF
contributing to \Eq{GF-exp} are calculated as
\be {\rm Res}_n=
\frac{\pi}{2i}\sqrt{\rho\rho'}J_m(n_r k\rho_<)J_m(n_r k\rho_>)r_m(k_n)  \label{Res2} \ee
with $r_m(k_n)$ found in \Eq{Res}. Given that the spatial dependence of the GF, as described by
Eqs.\,(\ref{GF-exp}), (\ref{Res2}), and (\ref{deltaG}), is represented by products of the RS wave
functions $R_m(\rho,k_n)$ and their analytic continuations $R_m(\rho,k)$ with $k$-values taken on
the cut, we arrive at the GF in the form of Eqs.(\ref{Gm}) and (\ref{Sigma}) which are then used in the RSE, once the
modes have been converted into $\Enb$ form using the analytics in appendix A.

\section{Homogeneous cylinder perturbation}

In this appendix is given the perturbation matrix elements for the homogeneous perturbation of the TE modes of
a homogeneous cylinder. I give the analytic expression for the perturbation matrix elements required to produce the
numerical results in \Sec{sec:Hom}, numerical results which validate the analytics which I developed for this manuscript.

The homogeneous perturbation \Eq{eps-hom} does not mix different angular momentum $m$-values. The matrix elements
between RS with the same azimuthal number $m$ are given by the radial overlap integrals
\begin{multline}\label{H1}
V_{\bar{n}\bar{n}'}=\dfrac{\hat{A}^{\rm TE}(k_{\bar{n}})\hat{A}^{\rm TE}(k_{\bar{n}'})\gamma(k_{\bar{n}})\gamma(k_{\bar{n}'})\Delta\epsilon}{2}
\\
\times\left[I_{m-1}(k_{\bar{n}},k_{\bar{n}'})+I_{m+1}(k_{\bar{n}},k_{\bar{n}'})\right]\, 
\end{multline}
where,
\be \gamma( k_{\bar{n}})=\left\{
\begin{array}{cl}
\sqrt{|m|(n^2_r-1)} & \text{for\ \  } k_{\bar{n}}=0\,,\\
1 &\text{otherwise\ \  } 
\end{array} \right.
\ee
and
\be
I_{j}(k_{\bar{n}},k_{\bar{n}'})=\int_RJ_{j}(nk_{\bar{n}}\rho')J_{j}(nk_{\bar{n}'}\rho')\rho'd\rho' \ee
yielding for identical basis states ($\bar{n}=\bar{n}'$)
\begin{multline}
I_{m}(k_{\bar{n}},k_{\bar{n}'})\\
=\frac{R^2}{2}\left[[J_{m}(n_r k_{\bar{n}}R)]^2-J_{m-1}(n_rk_{\bar{n}}R)J_{m+1}(n_r k_{\bar{n}}R)\right] \label{V-homogeneous1}
\end{multline}
and for different basis states ($\bar{n}\neq\bar{n}'$)
\begin{multline}
\dfrac{R}{k^2_{\bar{n}}-k^2_{\bar{n}'}}\cdot\dfrac{k_{\bar{n}'}}{n_r}J_{m}(k_{\bar{n}}n_rR)J_{m-1}(k_{\bar{n}'}n_rR)
\\
-\dfrac{R}{k^2_{\bar{n}}-k^2_{\bar{n}'}}\cdot\dfrac{k_{\bar{n}}}{n_r}J_{m}(k_{\bar{n}'}n_rR)J_{m-1}( k_{\bar{n}}n_rR)
\\
=I_{m}(k_{\bar{n}},k_{\bar{n}'})
\end{multline}

When $k_{\bar{n}}=0$ we can use the asymptotic form of the Bessel function,
\be
J_m(z)=\dfrac{(z/2)^m}{\Gamma(m+1)}
\ee
\be
\Gamma(m+1)=m!
\label{H6}\ee
from which we see $k_{\bar{n}}\rightarrow 0$ cancels out everywhere in $V_{\bar{n}\bar{n}'}$.

There is an important transformation to note,
\be
\dfrac{{A}^{\rm TE}(k_{\bar{n}})}{J_m(k_{\bar{n}}n_rR)}=\hat{A}^{\rm TE}(k_{\bar{n}})
\ee

Hence with the analytic integrals in this appendix we are able to completely reproduce the analytic results in \Sec{sec:Hom}.

\section{Evaluating Equation A26}

In this appendix I analytically evaluate the important integral equation \Eq{HELIOS}
\begin{multline}\label{HELIOS2}
\dfrac{2{M^2_{k_{\bar{n}}}}}{n^4_r}=
\\
=\dfrac{\int\left[J^2_{m-1}(n_rk_{\bar{n}}\rho')+J^2_{m+1}(n_rk_{\bar{n}}\rho')\right]\rho'd\rho'}{R_{-}J_m(n_rk_{\bar{n}}R_{-})\dfrac{\partial{J_m(n_rk_{\bar{n}}R_{-})}}{\partial{r}}+k^2_{\bar{n}}\varepsilon\int J^2_{m}(nk_{\bar{n}}\rho')\rho'd\rho'}
\\
=\dfrac{2}{n^2_rk^2_{\bar{n}}}
\end{multline}

In order to evaluate \Eq{HELIOS2} let us consider the numerator in the above integral, taking into account standard integrals,
\begin{multline}\label{HELIOS2}
2\int^{R}_{0}\left[J^2_{m-1}(n_rk_{\bar{n}}\rho')+J^2_{m+1}(n_rk_{\bar{n}}\rho')\right]\rho'd\rho'
\\
=\left(R^2-\left(\dfrac{m-1}{k_nn_r}\right)^2\right)J^2_{m-1}(n_rk_{\bar{n}}R)+R^2\left[J_{m-1}'(n_rk_{\bar{n}}R)\right]^2
\\
+\left(R^2-\left(\dfrac{m+1}{k_nn_r}\right)^2\right)J^2_{m-1}(n_rk_{\bar{n}}R)+R^2\left[J_{m+1}'(n_rk_{\bar{n}}R)\right]^2
\end{multline}
with rearrangement and use of the following identities
\be\label{MMM}
J_{m-1}(z)-J_{m+1}(z)=2{J}_m'(z)
\ee
\be\label{ZZZ}
J_{m-1}(z)+J_{m+1}(z)=\dfrac{2m}{z}{J}_m(z)
\ee
we arrive at,
\begin{multline}
\int^{R}_{0}\left[J^2_{m-1}(n_rk_{\bar{n}}\rho')+J^2_{m+1}(n_rk_{\bar{n}}\rho')\right]\rho'd\rho'
\\
=\dfrac{R^2}{2}\left[J^2_{m-1}(n_rk_{\bar{n}}R)+J^2_{m+1}(n_rk_{\bar{n}}R)\right]
\\
-\dfrac{R^2J_m(n_rk_{\bar{n}}R)}{n_rk_{\bar{n}}R}\left[\left(m-1\right)J_{m-1}(n_rk_{\bar{n}}R)+\left(m+1\right)J_{m+1}(n_rk_{\bar{n}}R)\right]
\\
+J^2_{m}(n_rk_{\bar{n}}R)R^2
\end{multline}
Further use of the identities \Eq{MMM}, \Eq{ZZZ} and rearrangement gives
\begin{multline}
\int^{R}_{0}\left[J^2_{m-1}(n_rk_{\bar{n}}\rho')+J^2_{m+1}(n_rk_{\bar{n}}\rho')\right]\rho'd\rho'
\\
=\dfrac{R^2}{2m}\left[J^2_{m-1}(n_rk_{\bar{n}}R)-J^2_{m+1}(n_rk_{\bar{n}}R)\right]
\\
-R^2J_{m+1}(n_rk_{\bar{n}}R)J_{m-1}(n_rk_{\bar{n}}R)
\\
+\dfrac{(n_rk_{\bar{n}}R)^2}{4m^2}\left[J_{m-1}(n_rk_{\bar{n}}R)+J_{m+1}(n_rk_{\bar{n}}R)\right]^2
\end{multline}

Next we move on to the denominator
\begin{multline}
={RJ_m(n_rk_{\bar{n}}R)\dfrac{\partial{J_m(n_rk_{\bar{n}}R)}}{\partial{\rho}}+k^2_{\bar{n}}\varepsilon\int^R_0 J^2_{m}(nk_{\bar{n}}\rho')\rho'd\rho'}
\\
=\dfrac{(n_rk_{\bar{n}}R)^2}{4m}\left[J^2_{m-1}(n_rk_{\bar{n}}R)-J^2_{m+1}(n_rk_{\bar{n}}R)\right]
\\
+\dfrac{(n_rk_{\bar{n}}R)^4}{8m^2}\left[J_{m-1}(n_rk_{\bar{n}}R)+J_{m+1}(n_rk_{\bar{n}}R)\right]^2
\\
-\dfrac{(n_rk_{\bar{n}}R)^2}{2}J_{m-1}(n_rk_{\bar{n}}R)J_{m+1}(n_rk_{\bar{n}}R)
\end{multline}
where again I have made use of standard integrals, the identities \Eq{MMM}, \Eq{ZZZ} and rearrangement.

Hence making use of these evaluations we arrive at the important result,
\begin{multline}\label{HELIOS24}
\dfrac{2{M^2_{k_{\bar{n}}}}}{n^4_r}=
\\
=\dfrac{\int\left[J^2_{m-1}(n_rk_{\bar{n}}\rho')+J^2_{m+1}(n_rk_{\bar{n}}\rho')\right]\rho'd\rho'}{R_{-}J_m(n_rk_{\bar{n}}R_{-})\dfrac{\partial{J_m(n_rk_{\bar{n}}R_{-})}}{\partial{r}}+k^2_{\bar{n}}\varepsilon\int J^2_{m}(nk_{\bar{n}}\rho')\rho'd\rho'}
\\
=\dfrac{2}{n^2_rk^2_{\bar{n}}}
\end{multline}
or
\be
M^2_{k_{\bar{n}}}=\dfrac{n^2_r}{k^2_{\bar{n}}}
\ee
Where $M^2_{k_{\bar{n}}}$ is the factor that links $\Enb$ to $\Hnb$ in \Eq{LENNON}.

\section{Rotating the cut}

In this section I explain how to find the correct sheet of the Hankel function, this corresponds to rotating the cut in the
Hankel function on the complex plane. The correct sheet of the Hankel function gives purely outgoing plane cylindrical waves, infinitely far from the 
resonator.

First of all let us consider how the Hankel function is defined. The Hankel function can be expressed as \cite{Gradshtein}
\be H_m(z)=J_m(z)+ iN_m(z)
\label{Hankel} \ee
using a multiple-valued Neumann function
\be N_m(z)=\tilde{N}_m(z)+ \frac{2}{\pi}J_m(z) \ln\frac{z}{2} \label{Neumann} \ee
where $\tilde{N}_m(z)=z^m F_m(z^2)$ is a single-valued polynomial \cite{Gradshtein} while
$\ln z$ is a multiple-valued function defined on an infinite number of Riemann sheets.

However considering \Eq{Neumann}, since we know,
\be
\ln\frac{z}{2}=\ln\frac{ze^{i2\pi M}}{2}=\ln\frac{z}{2}+2\pi Mi
\ee
where $M$ is an integer, we can also say that,
\be 
H_m(z)=\left[J_m(z)+ iN_m(z)\right]-4MJ_m(z)
\label{Hankel2} 
\ee
Hence by \Eq{Hankel2} we can add any multiple of $4J_m(z)$ to $H_m(z)$ and still obtain a valid
solution to the following defining equation
\be\label{GTGT}
\left[\dfrac{\partial^2}{\partial\rho^2}+\dfrac{1}{\rho}\dfrac{\partial}{\partial\rho}-\dfrac{m^2}{\rho^2}+k^2\right]H_m(\rho k)=0
\ee

Let us investigate the meaning of this result further. Since
\be\label{FFFF}
\lim_{\rho\rightarrow\infty}J_m(\rho k)={\sqrt{\dfrac{2}{\pi k\rho}}}\left({e^{i\rho k}+e^{-i\rho k}}\right)
\ee
we can see from \Eq{Hankel2} that changing the value of the integer $M$ corresponds to altering the boundary condition for the solution of \Eq{GTGT}
such that as $M\rightarrow\infty$ we have an equal balance between incoming and outgoing waves, infinitely far from the cylinder.

Hence in order to find the Hankel function with the correct boundary conditions we must consider the general solution to \Eq{GTGT} in the limit
of $\rho\rightarrow\infty$, where \Eq{GTGT} becomes
\be\label{GTGT2}
\left[\dfrac{\partial^2}{\partial\rho^2}+\dfrac{1}{\rho}\dfrac{\partial}{\partial\rho}+k^2\right]H_m(\rho k)=0
\ee
and
\be\label{FFFF2}
\lim_{\rho\rightarrow\infty}H_m(\rho k)={\sqrt{\dfrac{2}{\pi k\rho}}}\left({Ae^{-i\rho k}+Be^{+i\rho k}}\right)
\ee
is the sum of incoming and outgoing plane waves.

We require in \Eq{FFFF2} $A=0$. We achieve this by first calculating the ratio $A/B$ and then adding or subtracting multiples of $4J_m(\rho k)$
to $H_m(\rho k)$ in order to obtain a solution with the correct boundary conditions.

The calculation of $A/B$ is as follows,
\be\label{kurt_cobain}
\lim_{\rho\rightarrow\infty}\dfrac{H_m(-\rho k_n)}{H_m(+\rho k_n)}=\dfrac{Ae^{+i\rho k_n}+Be^{-i\rho k_n}}{Ae^{-i\rho k_n}+Be^{+i\rho k_n}}=\dfrac{A}{B}
\ee
noting $\Im(k_n)\leq 0$. If $A/B=0$ we have the correct sheet of the Hankel function. The \Eq{kurt_cobain} should be evaluated numerically on the computer
for large values of $\rho$, convergence will be quick due to the exponential nature of the functions involved.

Hence I have formulated a numerical recipe to rotate the cut and find the correct sheet of the Hankel function.
The correct sheet of the Hankel function is the sheet which provides us with outgoing boundary condition for use in the
calculation of the RS eigenmodes (see pairs \Eq{secular2} and \Eq{Dfunc1}, \Eq{secular3} and \Eq{Dfunc2}).

\end{document}